\newcommand{\beq}{\begin{quote}}
\newcommand{\enq}{\end{quote}}
\newcommand{\be}{\begin{eqnarray}}
\newcommand{\en}{\end{eqnarray}}
\newcommand{\del}{\delta}

\documentclass[twocolumn,showpacs,preprintnumbers,amsmath,amssymb]{revtex4}

\usepackage{graphicx}
\usepackage{dcolumn}
\usepackage{bm}

\begin{document}

\title{ A critique of q-entropy for thermal statistics 
} 
\date{November 10}
\author{Michael Nauenberg\\
Department of Physics\\
University of California, Santa Cruz, CA 95064 
}
\email{michael@mike.ucsc.edu}

\begin{abstract}
During the past dozen years there have been numerous  articles 
on a relation between entropy and probability
which is non-additive and has a 
parameter $q$ that depends on the nature of the thermodynamic 
system under consideration.
For $q=1$ this relation corresponds to the Boltzmann-Gibbs
entropy, but for other values of $q$  it is claimed that 
it leads to a formalism which is consistent 
with the laws of thermodynamics.
However, it is shown here that the joint entropy 
for systems having {\it different} values of
$q$ is not defined in this formalism, 
and consequently fundamental thermodynamic concepts 
such as temperature and heat exchange cannot be considered 
for such systems.
Moreover, for $q\ne 1$  the probability distribution for weakly interacting
systems  does not factor into the product of the probability
distribution for the separate systems,  leading to spurious correlations
and other  unphysical consequences, e.g. non-extensive
energy, that  have  been ignored in various applications
given in the literature.

\end{abstract}


\maketitle

\subsection*{Introduction}
In 1988 a relation between entropy  and probability 
for thermal statistics
was  proposed by C. Tsallis  \cite{tsallis1} which
is non-additive  and depends on a parameter $q$ that is presumably  
determined by the nature of the  thermodynamic system under consideration. 
For the special case  $q=1$ this relation 
reduces  to the Boltzmann-Gibbs entropy, but for other values
of $q$ it is claimed that this  q-entropy
leads to  an  alternative  formalism that is ``entirely consistent''  
with the laws of thermodynamics
\cite{tsallis1,curado1,tsallis2,tsallis3,baranger1,cohen1}. 
It will be shown, however, that this claim is not valid, because
the total entropy of  thermodynamic  systems 
that are in thermal contact
is not even  defined in this formalism 
for the case where systems have q-entropies 
with {\it different} values of $q$. 
As a consequence, the fundamental concepts of temperature
and heat exchange between  such  systems  can  not be introduced
in this  formulation of thermodynamics unless $q$ is a universal constant.
But even in this case, maximizing the q-entropy of the combined systems
leads to unphysical results unless q=1.  
In particular, for $q\ne 1$  the resulting 
joint probability for  states of weakly coupled systems 
is not the product of the individual probabilities,  
and as a consequence  the total  energy of the combined systems is
not the sum of the mean energies of each  system,  although
the additivity  property is assumed to be satisfied by the individual  
micro-states. 
This   non-additive property of the mean energy in the q-entropy
formalism is manifestly incorrect, and not surprisingly 
leads to unphysical results, 
but the consequences  have been ignored
in various applications of q-entropy in the literature, 
as will be illustrated in some of the  examples discussed here.

The  definition of q-entropy for a thermodynamic  system with micro-states
labelled by an  index $i$ is given by \cite {tsallis1,curado1,tsallis2,
tsallis3} 
\be
\label{entrop1}
S_q=k\frac{(1-\sum_i p_i^q)}{(q-1)},
\en
where $k$ is a constant, $q$ is an undetermined  parameter, 
and the quantities $p_i$ are positive 
numbers which satisfy  the condition
$\sum_i p_i= 1$.  In the limit that  $q\rightarrow 1$ one recovers  
the Boltzmann-Gibbs form of the  entropy 
\be
\label{bgentropy}
S_1=-k_B\sum_i p_iln(p_i), 
\en
where $k=k_B$ is the Boltzmann constant.
In this special case, $p_i $ is the probability 
for the occurrence of the $i$-th micro-state, and  
this identification has been extended 
to the case $q\ne 1$. This extension, however, is not valid  
as can be seen  from the definition 
of  mean values for  physical quantities associated with
the q-entropy. For example, the internal energy $U_q$ 
is given by the form
\be
\label{ener1}
U_q=\sum P_i \epsilon_i,
\en 
where $\epsilon_i$ corresponds to the $i$-th energy eigenvalue 
of the system, and $P_i=p_i^q/\sum_j p_j^q$.
The  quantities $P_i$ are  called ``escort'' probabilities
in the q-entropy literature, 
but according to the conventional definition of 
mean value in statistics, these $P_i's$ are 
the actual probabilities for the states of the system.
Hence for $q\ne 1$  the 
$p_i's $ introduced in the definition of q-entropy, Eq. \ref{entrop1}, 
are devoid of any  physical
meaning, and are just functions
of the probabilities  $P_j$ according to the relation
\be
\label{prob1}
p_i=P_i^{1/q}/(\sum_j P_j^{1/q}).
\en
  
For values of $q\ne 1$, the 
q-entropy expression introduced in  Eq. \ref {entrop1} is shown to be non-additive by the
following arguments \cite{tsallis1}-\cite {tsallis3}. 
Suppose that two thermodynamic  systems $A$ and $B$ 
are weakly coupled or are the subsystems of a larger system,  
and {\it assume} that the joint probabilities
for the states of the combined system are 
the product of the probabilities for the states
of the individual systems. As we shall see later on,
this fundamental factorization property is not satisfied
by the q-entropy formalism, but surprisingly this  fact
has  been ignored in the literature. According  
to Eh. \ref{prob1}, we then have  
\be
\label{probAB}
p_{i,j}^{AB}=p_i^A p_j^{B}.
\en
Substituting this form into the expression for 
q-entropy,  Eq. \ref{entrop1}, one  obtains the relation 
\be
\label{entrops1}
S_q(AB)=S_q(A)+S_q(B)+(1-q)S_q(A)S_q(B)/k.
\en
However, this non-additive relation for the q-entropy leads 
immediately to a difficulty in the interpretation of $S_q$ as an
expression for the thermodynamic  entropy \cite {guerb1}. 
Since weak coupling means that  the energy eigenvalues of the combined systems
are essentially additive, we have
\be
\label{eAB}
\epsilon_{i,j}^{AB}=\epsilon_i^A+\epsilon_j^B,
\en
and, according to Eqs.\ref{ener1} and \ref{probAB}, the 
total mean energy $U_q(AB)$ of the combined system  
is also  additive:
\be
\label{energyt}
U_q(AB)=U_q(A)+U_q(B).
\en
Assume now that the combined system is isolated while there is an 
infinitesimal exchange of energy between
systems $A$ and $B$.
Then the variations $\delta S_q(AB)=0$  and $\delta U_q(AB)=0$, 
which implies that 
\be
\frac {\delta S_q(A)}{1+(1-q)S_q(A)/k}= -\frac{\delta S_q(B)}{1+(1-q)S_q(B)/k},
\en
and
\be
\label{energy1}
\delta U_q(A)=-\delta U_q(B).
\en
Combining these two equations, one finds that
\be
\label{temp1}
[1+(1-q)S_q(A)/k]T(A)= [1+(1-q)S_q(B)/k]T(B),
\en
where $T(A)$ and $T(B)$ are the 
absolute temperatures defined by the standard 
thermodynamic  relation
\be
\frac{\partial S}{\partial U}=\frac{1}{T}.
\en
For two systems in thermal contact these two temperatures should  
be equal, but according to Eq.\ref {temp1} this applies  only if $q=1$. 
This problem is not unexpected, because temperature should be an
an {\it intensive} 
quantity, but this is not possible in a  formalism 
where the energy is additive, Eq.\ref{energy1}, while the entropy does not 
satisfy this property.
To avoid this problem, it has been proposed \cite{tsallis3,martinez2, casas1} 
to re-define absolute temperature as the quantity
\be
\label{newtemp}
T_q=(1+(1-q)S_q/k)T
\en
in which case  the condition for thermal equilibrium,  $T_q(A)=T_q(B)$,
is satisfied by Eq. \ref{temp1}.
But this definition of temperature, which must be universal, 
cannot be extended to  
systems $A$ and $B$ which are described by  q-entropies 
with {\it different} values  $q_A$ and $q_B$. 
In this case the 
q-entropy of the combined system, which is characterized
by the quantities $p_{i,j}^{AB}$, Eq. \ref{probAB}, is not defined
in term of these two $q$ parameters.
Following the q-entropy formalism, one would have to introduce a 
new parameter $q'$ for the the q-entropy and energy  of the
combined system, but then these  thermodynamic  variables  cannot
be expanded in terms of the corresponding variables  for the component
systems $A$ and $B$ \cite {sasaki1},
\footnote{ In reference \cite{sasaki1} the authors define the  entropy 
                    of the joint system  $AB$ in terms 
                    of an {\it ad-hoc} function of the q-entropies
                    $S_{q_A}(A)$ and $S_{q_B}(B)$ of systems $A$ and $B$. 
                    This procedure, however, ignores
                    the consistency requirement
                    that this joint system must also  
                    be described by the same  q-entropy relation, 
                    Eq.\ref {entrop1}}.
Instead, according to Eq.\ref{entrops1}, the q-entropy of the combined 
system  would be given in terms of {\it pseudo} q-entropies for systems  $A$ and $B$ with
the {\it same} new parameter $q'$, and  a similar  problem would occur with
the expansion of the total energy, Eq.\ref{energyt}, i.e.
\be
S_{q'}(AB)=S_{q'}(A)+S_{q'}(B)-(1-q')S_{q'}(A)S_{q'}(B)/k,
\en
and
\be
U_{q'}(AB)=U_{q'}(A)+U_{q'}(B).
\en
Consequently, the concepts of thermal equilibrium, temperature,
and heat exchange cannot  be formulated for such systems.
For example, no meaning can be attached to the statement
that a system described by the Boltzmann-Gibbs entropy  is 
in thermal equilibrium with a system described by  q-entropy
with $q\ne 1$.   In other words, a Boltzmann-Gibbs 
thermometer would not be able to measure the temperature of
a  q-entropic system, and the laws of thermodynamics 
would therefore fail to have general  validity
\footnote{In reference \cite {tsallis3}, Tsallis states  that
                    `` if we have in thermal contact systems with 
                    different entropic indices, say $q_A$ and $q_B$,
                    it seems {\it plausible} [my italics] that at equilibrium
                    $ T_{q_A}(A)=T_{q_B}(B)$.''
                    But as we have shown here,  such a relation 
                    is not justified by the q-entropy formalism.}.

It follows that the parameter $q$ 
must be a {\it universal} constant, just like the
Boltzmann constant $k$, which is 
applicable to {\it all} systems in thermodynamic  equilibrium.
If q is universal, a thermodynamic formulation for an infinitesimal  reversible 
transfer of heat $dQ$ can be given between systems $A$ and $B$,  
with
\be
dQ=T(A)dS_q(A)=-T(B)dS_q(B),
\en
corresponding to an exchange $dU_q (A)= -dU_q (B)$
in the internal energy of these systems. 
But one is  faced with the problem that 
in this case the temperature 
$T(A)$ is not equal to $T(B)$, which violates a fundamental
principle  of thermodynamics for systems in thermal equilibrium.
Moreover,  the  corresponding differentials  $T_q(A) dS_q(A)$ and 
$T_q(B)dS_q(B)$ associated with the proposed
re-definition of absolute temperature, Eq.\ref{newtemp}, 
do not have any physical significance.  
In principle, this problem can be solved \cite{abe3}
by  introducing a different form $S_q^R$ for the q-entropy  
in the  thermodynamic relation 
for the temperature $T_q$, so that
\be
\frac{\partial S_q^R}{\partial U_q}=\frac{1}{T_q}.
\en
From the definition of $T_q$ given by Eq. \ref{newtemp} it follows that
\be
\label{rey1}
S_q^R=\frac{k}{(1-q)}ln(1+(1-q)S_q),
\en
This alternative expression for the q-entropy 
was introduced by A. R\'eyni \cite{reyni1} in the form
\be
\label{rey2}
S_q^R=\frac{k}{(1-q)}ln(\sum_i p_i^q),
\en
which  is additive 
\footnote { In reference \cite{abe3}
                    the author concludes  ``that at the
                    level of macroscopic thermodynamics, transmutation
                    occurs from Tsallis theory  
                    to R\'eyni-entropy based theory''. 
                    This appears to be a  convoluted admission that 
                    a non-additive form
                    of q-entropy such as  Eq. \ref{entrop1} 
                    is incompatible with the laws of thermodynamics}, 
as can be verified  by  substituting 
for the $p_i's$ the form in Eq. \ref{probAB}. 

However, there are additional problems with either of these two
definitions of q-entropy, even
when $q_A=q_B=q$. Maximizing such entropy functions subject to the constraint 
of constant energy, Eq. \ref{ener1} yields the q-probability distribution
\cite{tsallis1,curado1,tsallis2,tsallis3}
\be
\label{prob2}
p_i \propto [1-(1-q)\beta \del \epsilon_i]^{1/(1-q)},
\en
where $\del \epsilon_i=\epsilon_i-U_q$,
$U_q$is the mean energy, Eq. \ref{ener1}, and $\beta$ is a parameter  
related to the inverse temperature.
It is clear that  the corresponding distribution $p_{i,j}^{AB}$ 
for the combined system $AB$ does 
not factor into the product $p_i^Ap_j^B$ even when 
the energies of the micro-states  are additive, Eq. \ref{eAB}, 
unless $q=1$. For example,  to first order in $q-1$
\footnote{A similar expression
is obtained in \cite{casas1} which is, however, not quite correct,
because in the second term a factor $p(\epsilon_i^A)p(\epsilon_j^B)$
is left out } one finds that 
\be
p(\epsilon_i^A+\epsilon_j^B)=p(\epsilon_i^A)p(\epsilon_j^B)[1+(q-1)
\beta^2(\del \epsilon_i^A \del \epsilon_j^B)]
\en
In the  limit $q=1$  Eq. \ref{prob2} reduces 
to the Boltzmann-Gibbs exponential 
form for the probability \cite {gibbs1},
\be
\label{pBG}
p_i \propto exp(-\epsilon_i/kT),
\en
where $\beta=1/kT$.
As is well known, this canonical distribution
follows {\it uniquely} from the factorization requirement
that 
\be
p(\epsilon_i^A)p(\epsilon_j^B)=p(\epsilon_i^A+\epsilon_j^B). 
\en
The  Boltzmann-Gibbs form for the entropy $S$, Eq. \ref{bgentropy}, then follows 
from the assumption that 
\be
S=\sum_i f(p_i).
\en
where the function  $f(p)$ is determined uniquely.
Applying  the thermodynamic definition of temperature, 
$\partial S/\partial U =1/T$ and Boltzmann-Gibbs definition
of the  probability distribution, Eq. \ref{pBG}, one obtains the relation
\be
\sum_i p_i(\epsilon_i-U)[\frac{df(p_i)}{dp_i}+kln(p_i)]=0.
\en
Hence 
\be
f(p_i)=-kp_iln(p_i)
\en
provided $f(1)=f(0)=0$, which corresponds to the
requirement that the entropy vanishes  at $T=0$.

\subsection*{Unphysical  properties  resulting from applications 
of q-entropy to thermodynamic systems}

To illustrate the consequences of disregarding such basic  
considerations, I would like to call attention to some  
unphysical results, left unmentioned in the literature,
that follow from recent applications of q-entropy to some well-known 
thermodynamic systems.
For example, many papers have been published 
on the application of q-entropy to black-body radiation 
\cite {black1,black2,black3,black4,black5,black6,black7,black8}. 
After laborious analysis and  
numerical computations, the authors in references \cite
{black5,black6,black7} find that 
for $q\ne 1$ there are  deviations from the well-known Stefan-Boltzmann law
which states that the  radiation energy depends on the 
fourth power of the temperature
\footnote { The calculations in references \cite{black1,black2,black3,black4,black5}
            do not exhibit  deviations from the Stefan-Boltzmann law  
	    because these calculations have left out an anomalous 
            dependence on  $V/V(T)$, where $V$ is the volume of the
            cavity and $V(T)=(\hbar c/kT)^3$.  It is clear, 
            that such a dependence  must appear, because both the energy
            and the  entropy density are non-extensive  for $q \ne 1$ (except to
            lowest order in an expansion in $q-1$, in which case  the energy is
            extensive).}. 
But since Boltzmann derived
this result from purely thermodynamical reasoning, without any 
statistical assumptions about the form
of the entropy, it seems at first sight strange that 
such a deviation can occur in a formalism  which is
supposed to satisfy the laws of thermodynamics.  
The explanation  is that as a consequence of
the non-extensivity property of the q-entropy formalism,
the black-body energy as well as the entropy  
do not dependent linearly on the volume of the cavity,
as was originally assumed by Boltzmann. 
Hence, the {\it q-energy density} and the {\it q-entropy density} depend 
on the volume $V$ of the cavity, although there is no comment concerning this unphysical
property  in any of the q-entropy calculations
in references \cite {black1,black2,black3,black4,black5,black6,black7,black8}.   
As follows from simple dimensional arguments, 
the volume dependence must be given
by a dimensionless parameter $V/a(T)^3$, where $a(T)$ is a characteristic
length which can depend only on the temperature $T$. From statistical mechanics we
learn that $a(T)=(\hbar c/kT)$ in fact corresponds to the
mean thermal wavelength of the black-body photons. Moreover, 
expansions in a power series 
of this parameter which have been  applied to fit  the cosmic background radiation
\cite{black1,black2,black3,black4,black5,black6,black7,black8}
are nonsensical,because in this case $a(T)$ is of order $1/10$ $cm$
while the cavity volume $V$ has cosmological  dimensions!
On purely thermodynamic grounds it can also  be shown that if the temperature  dependence of
the black-body energy density were to have  the form  
$u\propto T^{4+\delta}$, 
then the Maxwell-Boltzmann relation $p=(1/3)u$ for
the thermal radiation pressure $p$ would lead to
a power law volume dependence 
$u\propto V^{\delta/3}$, and correspondingly  $U \propto T(T^3V)^{1+\delta/3}$,
in accordance with our previous dimensional argument. Similarly, one finds that 
$s\propto V^{\delta/3}$ for the entropy density.
But unless $\delta=0$, such a volume dependence is incompatible with 
Kirchhoff's law which states that the ratio of emissivity to absorption
of radiation in the walls of the cavity must be a universal function of the 
temperature, and the frequency of the radiation.  These properties are
required in order that such a  cavity reach thermal equilibrium.
Historically, this law  was the original  basis for  
the universality  properties
of  black-body radiation that culminated in Planck's 
famous derivation.

Actually, the application of the q-entropy formalism  to any system 
which consists of {\it non-interacting} components leads to  
unphysical properties. For example, for an ideal gas the 
energy is found to depend nonlinearly on the number of particles,
the gas pressure $p$ is not equal to  $2/3$ of the energy
density, and there are correlations between the energies
of any pair of particles\cite{ideal1, ideal2,ideal3},    
contrary to very well-known  result of  kinetic gas theory
and statistical mechanics. Similarly,  for  a system of 
non-interacting magnetic moments  anomalous dependences 
are found for the magnetization and
susceptibility on the number of spins \cite {spin1}, 
although  claims have been also made that the magnetization 
reproduces experimental results in certain manganites 
\cite {spin2,spin3}
The reason for the failure of this formalism to give physically sensible results
is not hard to see - it is due to the fact that 
the  probability distribution,
Eq. \ref{prob2}, for any two components $A_1$  and $A_2$ is  
not the product of the 
probabilities for  the separate components as would be expected if these
components are {\it non-interacting}. Hence, contrary to basic physical
principles, this formalism  gives rise to spurious correlations among  
these components of such systems.  

Some proponents of  q-entropy have  argued that this formalism  should be 
considered  only for systems for which the Boltzmann-Gibbs 
thermodynamic formalism supposedly ``fails''. 
Frequently mentioned as candidates  are systems 
with components that interact primarily through  long range forces    
such as  gravitational forces  for which  the total  energy and entropy
are  non-extensive.  In Nature, such system  correspond to 
astrophysical objects such as stars and
galaxies, but these objects are generally not found 
at {\it maximum entropy} and (correspondingly)
{\it uniform} temperature.
For example,  stars  are either evolving slowly  in time, 
like main sequence stars, which  have a large temperature gradient
from the interior to the surface, and emit thermal radiation
(black-body energy and  entropy), or else  have reached a degenerate 
state  such as  dwarf stars  or neutron stars  provided 
certain mass limits are satisfied \cite {chapline}. 
Otherwise, stars eventually  
explode into supernovas, sometimes leaving remnants which
collapse into such  degenerate
states or into a black hole (which are states having  maximum
entropy).  
The equilibrium properties of stars are obtained  by 
hydrostatic equations  supplemented by {\it local} thermodynamic
equations for matter and radiation based on the Boltzmann-Gibbs entropy
\cite {weisskopf} \footnote{
According to Boltzmann-Gibbs statistics, 
a gas of particles interacting via gravitational
forces in thermal equilibrium (maximum entropy)  
would have to have infinite size and mass
For the solution of the isothermal gas polytrope see
\cite{chandra1}.}
Rather than being a "failure", the Boltzmann-Gibbs
statistics  has been applied to stellar structure  with enormous  
success.  A  q-entropy formalism, however, predicts the existence of finite
{\it isothermal} polytropes as the end products  of stellar evolution
( states of maximum entropy) which fails completely
to account for the observed property of stars in our universe.
Such q-polytrope solutions have also been discussed as models for galaxies 
\cite{plastinograv,taruya,chavanis}. 

Another example which  has been cited as a so-called  failure of 
Boltzmann-Gibbs entropy because it involves long-range 
electromagnetic forces,
is the divergence of the partition function $Z$ calculated for the
bound states of the hydrogen atom \cite {hydrogen1}.
In this case 
\be
Z=\sum_n exp(-\epsilon_n/T),
\en
where $\epsilon_n=-R/n^2$ are the
bound-state  energy levels, and $R$ is the Rydberg constant.
But the divergence of $Z$, which occurs in this case because 
the terms of the series approach unity
as $n$ becomes large,
is  related to  the growth of the 
the mean radius of the hydrogen atom which increases
as $n^2$. Obviously, in a gas of hydrogen  atoms
in thermal equilibrium, this radius can not become larger 
than the mean distance between atoms. Therefore 
this distance  provides an effective cutoff for the applicability
of the hydrogen  bound-state energy eigenvalues in the partition sum, 
because for larger values of $n$ 
these atoms can no longer  be treated even approximately as a gas 
of  non-interacting particles. Instead, for these states 
the gas must be viewed  as
a neutral plasma of electrons and protons 
interacting via long range electromagnetic forces. Hence
the q-entropy formalism, which supposedly  gives a finite partition 
function \cite {hydrogen1}, 
actually fails  to account for the correct physics 
of this problem. 
These and other failures \cite {neutrinos1}, 
\footnote{ It is frequently  mentioned \cite {tsallis2} 
                     that the {\it solar neutrino
                     problem} discussed in reference \cite{neutrinos1} 
                     is an example of  ``plain failures'' of
                     the Boltzmann-Gibbs entropy, because  the
                     calculated rates of nuclear reactions 
                     in the Sun supposedly  do not account for  
                     the neutrino observations. 
                     Recently, however, the SNO  data 
                     (Q.R. Ahmad et. al., Physical Review 
                     Letters 89 (2002)) show that there is now excellent 
                     agreement between theory and neutrino observations.  
                     Among other things, this result is a spectacular new 
                     confirmation of the  applicability
                     of Boltzmann-Gibbs entropy in the interior of a star}. 
in the application of q-entropy 
to well-known physical systems mirror the  
inconsistencies which are inherent  in a formulation of thermodynamics 
based on  q-entropy, Eq. \ref{entrop1}. 

\subsection{Conclusion}

We have shown that a {\it prerequisite}  to have a q-entropy formalism  
which is consistent with
the laws of  thermodynamics is that the parameter $q$
must be a {\it universal} constant, as is  the case also  with  
the  constant $k$  that corresponds to Boltzmann's constant for $q=1$. 
Moreover, for weakly coupled systems  the 
entropy as well as the energy must be additive, which is a condition  explicitly
violated by the Tsallis entropy, Eq. \ref{entrop1}. 
Although this condition  appears to be satisfied 
by the R\'eyni q-entropy, Eq. \ref{rey2}, this is actually  not the case, 
because the probability
distribution obtained by maximizing this entropy for $q\ne 1$, Eq. \ref{prob2}, 
does not satisfy the required  factorization condition, Eq. \ref{probAB},  
which is required for weakly coupled systems.
Thus, we have shown that the only value of $q$ consistent  
with the laws of thermodynamics 
is $q=1$, which corresponds to the familiar Boltzmann-Gibbs form for the  entropy,
Eq. \ref {bgentropy}.
Indeed, it has been explicitly demonstrated here that the application of
a q-entropy  formalism to  black-body radiation, and to  other systems with
weakly interacting components, leads to unphysical results when $q \ne 1$.

It has been suggested in the literature  that q-entropy calculations are
useful because they  provide  an additional parameter $q$
for comparing theory with observations, but this  rationale
fails to take into account the fact that such calculations would be
inconsistent with the fundamental principles of thermodynamics
and statistical mechanics.
Hence, if a small  departure of $q$ from unity is found
in a fit to data, as is  claimed for example
in various  analyses of the cosmic black-body radiation \cite {black1,black2,
black3,black4,black5,black6,black7,black8},
such a fit can not provide any physical insight whatsoever  
into the source or meaning of the deviations.

Finally, we remark that the  q-probability  distribution, Eq. \ref{prob2}, 
which is obtained by maximizing either of the two q-entropy relations, 
Eqs. \ref{entrop1} and \ref{rey2}, 
constrained  to a  fixed pseudo-energy function,
has also been applied to {\it non-equilibrium} problems. 
For example, data on turbulence in a pure-electron plasma column \cite{bogo1}
and the  velocity distribution in turbulent flow 
\cite {bodenschatz} \cite{swinney}
have  been fitted by  such a probability function \cite {beck1,beck2}.
But  absolutely no physical justification has been given 
for applying to non-equilibrium systems a fundamental condition -
maximum entropy - which is associated in statistical
mechanics with systems in thermal equilibrium. 
While such a probability distribution can also be obtained
from  other {\it ad-hoc}  assumptions \cite{beck1,beck2}, 
its connection  to  q-entropy and non-extensivity is completely  unfounded. 
Although  the q-probability distribution appears to be  a good phenomenological
parameterization for some turbulence data, its deduction  
from physical principles  has not been established  
\footnote { In references \cite{beck1,beck2} the author expressed the q-probability
 distribution, Eq. \ref{prob2}, as a Laplace integral, 
                     which is the basis for his suggestion that
                     this distribution is due to temperature fluctuations 
                     in ordinary  Boltzmann-Gibbs statistics. 
                     But no connection based on physics has been  made between these 
                     assumed temperature  fluctuations and q-entropy}. 
.

\subsection*{Acknowledgments}

I would like to thank S. Abe, C.Beck, E.G. Cohen,  A. Plastino, J. Riera and C. Tsallis for comments,
and for providing  references to the extensive  literature on non-extensive  q-entropy.

\end{document}